\documentclass[oribibl]{llncs}

\usepackage{url}
\usepackage{amsmath}
\usepackage{balance}
\usepackage{booktabs}
\usepackage{tabularx}
\usepackage{graphicx}
\usepackage{threeparttable}

\usepackage[colorlinks=true,linkcolor=cyan,
  citecolor=cyan,urlcolor=cyan]{hyperref}

\usepackage{fancyhdr} 
\fancypagestyle{firststyle}
{
   \fancyhf{}
   \fancyfoot[C]{\scriptsize{International Cybersecurity Law Review, published online in June 2022. The final publication is available at Springer via \url{https://doi.org/10.1365/s43439-022-00057-8}}}
}

\begin{document}

\title{A Review of Product Safety Regulations in the European Union}

\author{\normalfont Jukka Ruohonen \vspace{4pt}\\ University of Turku, Turku, Finland \\ \url{https://orcid.org/0000-0001-5147-3084}  \\ \texttt{juanruo@utu.fi}}


\maketitle

\begin{abstract}
Product safety has been a concern in Europe ever since the early 1960s. Despite
the long and relatively stable historical lineage of product safety regulations,
new technologies, changes in the world economy, and other major transformations
have in recent years brought product safety again to the forefront of policy
debates. As reforms are also underway, there is a motivation to review the
complex safety policy framework in the European Union (EU). Thus, building on
deliberative policy analysis and interpretative literature review, this paper
reviews the safety policy for non-food consumer products in the EU. The review
covers the historical background and the main laws, administration and
enforcement, standardization and harmonization, laws enacted for specific
products, notifications delivered by national safety authorities, recalls of
dangerous products, and the liability of these. Based on the review and analysis
of these themes and the associated literature, some current policy challenges
are further discussed.
\vspace{5pt}
\end{abstract}

\section{Introduction}

\thispagestyle{firststyle} 

Safety has reached a global priority in the face of the global COVID-19
crisis. Yet safety---understood in the present context as a risk to human
health---has long been on the agenda of legislators around the world, including
those in the European Union. Although global pandemics---from the Spanish flu
through the swine flu to the present crisis---have often captured the attention
in popular discourse, law-backed preparations for hazardous accidents have been
implemented from the 1960s onward. These preparations address also concerns that
range from terrorism, crime, and radicalization to traffic, pollution, and
environmental hazards, all of which may also include safety
consequences~\cite{Boustras20, Jore19}. Interestingly, many of the incidents
that originally prompted the preparations have been forgotten or buried to
history books. The production, storage, and transport of chemicals is a good
example: the global and European safety legislations for chemicals moved forward
through crises; dioxin in 1976, toxic oil syndrome in 1981, methyl isocyante in
1984, amerithrax in 2001, and so forth and so
on~\cite{DuarteDavidson14}. Consumer products do not cause such large-scale
accidents, but the safety consequences from these often affect more humans than
individual hazardous accidents. Regulation of safety requirements for consumer
products is also particularly difficult.

Product safety is a subset in the larger jurisprudence of consumer
protection. Throughout the world, the rationale builds upon the economic and
information asymmetries between producers and consumers; the latter are in a
weaker position with respect to the former, both with respect to bargaining
power and knowledge~\cite{Weitzenboeck15}. As the history of the automobile
industry vividly demonstrates, the incentives of producers to blindly pursue
profits have resulted in many dangerously defective or even hazardous products
that have put consumers' lives at risk~\cite{Surdam20}. Given that later on
safety became an important competitive benefit in the car
industry~\cite{Rausand09}, black and white perceptions should be avoided, as
always. Nevertheless, the general rationale of product safety regulations is to
protect the weaker party. A particular focus has historically been placed upon
vulnerable groups, such as children, disabled, and elderly~\cite{Mugica90}. This
protection rationale appeared on the legislative agenda already in the early
1960s both in Europe and the United States~\cite{Joerges10}. It did not take
long for it to appear also in the European Economic Community; the early safety
directives were passed in the early 1970s.

However, fragmentation has prevailed to the present day. Although perhaps not as
visibly as in some other areas of consumer protection, also European product
safety legislations have suffered from fragmentation and incoherence. A partial
explanation originates from the domain; it is difficult to legislate consumer
goods and services due to the pace of innovation and technological progress. But
another partial explanation stems from different cultures and historical
trajectories; the power struggles between the organized interests of consumers
and producers affected regulatory traditions differently in different member
states~\cite{Austgulen20}. A similar struggle between organized interests has
affected consumer law at the EU-level~\cite{Peltonen10}. A further partial
explanation can be found among producers; there have often been diverging or
even strictly competing interests between sectors and producers, and their
locations in different member states and regions~\cite{Dewitt99}. Likewise,
there have been occasional arguments that the EU's safety regulations are used
for protectionist objectives~\cite{Joerges10}, particularly against Chinese
products~\cite{Howells14}. Concerns such as environmental consequences have
intensified the struggles.

These challenges translate into research challenges; the EU's product safety
policy has always been notoriously complex and difficult to understand. This
complexity provides a motivation for the present short review on the key
legislations in Europe. But there is a further motivation: reforms are already
underway for European product safety legislations due to the rapid technological
progress and its impact upon the economy and consumers with it. Electronic
commerce, data, platforms, and other ingredients of contemporary economy have
also changed the incentives, externalities, and asymmetries between producers
and consumers~\cite{Howells20, Jin20}. Although product safety does not perhaps
weigh as much as security and privacy in these new circumstances, it is not
difficult to imagine also new safety risks arising from, say, artificial
intelligence, robotics, and cyber-physical systems, along many other
technological development trends. Indeed, many novel but wretched safety
incidents have already occurred and been cataloged for artificial intelligence
applications~\cite{McGregor21}. Of these incidents, the legacy of Elaine
Herzberg is perhaps the most memorable and saddest example; she was the first
pedestrian having been killed by a self-driving car.

\section{Concepts, Approach, and Data}

A few preliminary remarks are in order before the actual review: a few basic
concepts should be clarified, some words should be said about the approach to
the review, and something should be noted about the empirical data used
alongside the review for a few illustrative points.

\subsection{Concepts}

Safety, security, hazard, and risk---among many related concepts---are
domain-specific, debated, and generally ambiguous terms~\cite{Boustras20}. In
terms of information security, which is a subset of a larger security conundrum,
a risk is sometimes understood as a probability that an attack occurs; other
times it is more specifically seen as a conditional probability resulting from a
threat and a vulnerability. From this perspective, protection of (information)
security implies protection against intentional attacks, whereas safety is more
about unintentional lapses~\cite{Jore19}. Given the context of consumer
products, it is also useful to frame unintentional harms to those that have
consequences for the health and well-being of humans. This framing aligns with
the concept of hazard, which is often understood merely as a potential source of
harm~\cite{Rausand09}. Despite the differences, for illustrative purposes, the
basic information security concepts can be translated to the safety context: a
vulnerability could be a defect in a product that exposes a safety threat to
human health, such as, say, a suffocation, a strangulation, or a serious
electromagnetic disturbance. In this review, as well as in the EU regulations,
such safety threats of consumer products exclude social, psychological, and
related factors with potential health consequences. It is also important to
underline that the particular regulations considered exclude food products,
medicine and drugs, and occupational health risks, among other things.

Product safety is presumably the earliest case of a risk-based approach to
regulation in the EU. Since the 1990s safety regulations have relied on a
precautionary principle: dangers to health and environment should be taken into
account through systematic, scientifically based risk
analysis~\cite{Sciascia06}. Although definitions vary across domains, a
risk-based approach according to the European safety regulations is seen to
generally cover three dimensions: risk assessment, risk management, and risk
communication~\cite{Purnhagen13}. These vary from a product to another. For many
products---from chemicals to cosmetics, risk assessments may involve rigorous
laboratory testing. For some other products, including software products,
assessments range from the following of standards, documentation, and sound
engineering practices to quality controls and safety verification~\cite{Bujok17,
  Moore13, Playle11}. Risk management, likewise, varies across products. For
many products, including both tear-and-wear hardware products and software
products, life-cycle management is usually present; a product should be safe
throughout its intended life in the hands of consumers~\cite{Rausand09}. It
should be also stressed that risk management is not only strictly about safety;
for producers cost-benefit analysis is often present as well~\cite{Joerges10,
  Sciascia06}. In the European Union risk communication carries a particular
weight: whenever a risk is found from a product, producers should communicate
the risk to public authorities and consumers. With these clarifications of the
basic terminology in hand, the approach taken for the review can be briefly
elaborated.

\subsection{Approach}

The review approach taken follows the tradition of practice-oriented policy
analysis. Unfortunately---just like with safety, there are no commonly agreed
definitions for policy analysis. Roughly, policy analysis revolves around the
questions of what, how, and why governments do what they do, and what difference
does it make~\cite{Dye76}. In this short review the focus is on what they do
with a policy. This policy refers to a set of European legislations and
standards designed to ensure the safety of non-food consumer products. Safety in
itself is seen as the primary (but not necessarily the only) answer to the why
question.

Then, it seems reasonable to maintain that most practitioners of policy analysis
would agree that it: (a)~requires sensitivity to a given policy space, which is
neither limited to a particular polity nor a decision-making system; (b)~cannot
be separated from politics; and (c)~involves a practical motivation of careful
evaluation of problems and, whenever possible, different solutions to these. The
policy space for product safety is not limited to the EU's parliamentary
decision-making; national safety administrations and standardization
organizations---among others---possess considerable power in both shaping and
interpreting the overall safety policy. Such power leads to politics, which, in
the present context, is also shaped by politicians as well as the organized
interests of producers and consumers. Furthermore, in the EU not only are
legislations lobbied but interest group politics occur also in non-legislative
administrative institutions~\cite{Joosen21}. But in what follows, only a limited
focus is placed on agency and the intentions of political actors. The
identification of problems and bottlenecks satisfies the practical motivation,
although policy recommendations are kept to a minimum as reforms are already
underway in the EU.

Also the epistemological bases for policy analysis vary. At least
interpretative, narrative, normative, critical, historical, positivist,
evidence-based, and deliberative policy analysis frequently appear in the
literature. The last one suits the purposes of this review well. Formulated in
the early 2000s as a critical response to the distinctively positivist policy
studies at the time, deliberative policy analysis builds on three pillars:
interpretation, deliberation, and practice-orientation~\cite{Hajer03,
  LiWagenaar19}. Although a few descriptive statistics are presented, the
analysis is based on a qualitative interpretation of the main policy artifacts,
the safety legislations in the EU. In terms of deliberation and
practice-orientation, the intention is to cover the main arguments in the
historical and present policy debates, assessing the relative merits of these,
and giving a unique input through a synthesis.

Due to the deliberative approach, the focus is further on high-level issues and
trends instead of legal, technical, and other nitty-gritties of some particular
safety policies. In other words, the primary audience contains not only
researchers and academia but also policy-makers and others on the European
democractic fora. Because these political roundtables are in Europe, the review
also excludes comparisons with other, non-EU countries and international
arrangements.

At the same time, the approach taken is a literature review. Quantitative
reviews (such as meta-analysis) and protocol-based (such as systematic
literature reviews) approaches do not go well with policy analysis. The reasons
for this claim are many, ranging from a need to assess historical developments
to the requirement to cover both politics and policies. Thus, a traditional,
interpretative approach is followed with the literature; the goal is to develop
a comprehensive understanding and critical assessment of existing knowledge via
in-depth reading~\cite{Boell15, Torraco05}. This classical approach does not
mean that the collection of literature would have been unsystematic. Many
relevant databases were queried, among them ScienceDirect, HeinOnline,
Taylor-Francis Online, SpringerLink, Wiley Online Library, IEEE Xplore, and
SAGE~Journals.

\subsection{Data}

The few descriptive statistics presented are based on the EU's Community Rapid
Information System (RAPEX)~\cite{EU21a}. Established in the early 2000s, RAPEX
is a database for tracking notifications sent by national safety authorities
about dangerous consumer products, excluding food and pharmaceutical products
but including clothing, cosmetics, toys, electronic appliances, and many other
product types. It is administrated by the European Commission like many
analogous safety tracking systems, including those related medicines, drugs,
food products, serious cross-border health threats, and chemicals
incidents~\cite{DuarteDavidson14, Oxford14}. Although a database was established
already in the 1990s for alerts on dangerous consumer goods~\cite{Sanchez99}, it
is no longer publicly available online; RAPEX provides records from 2005
onward. In total, $n = 28,129$ entries were filed to it between 2015 and 19 January 2021.

\begin{figure}[th!b]
\centering
\includegraphics[width=8cm, height=4cm]{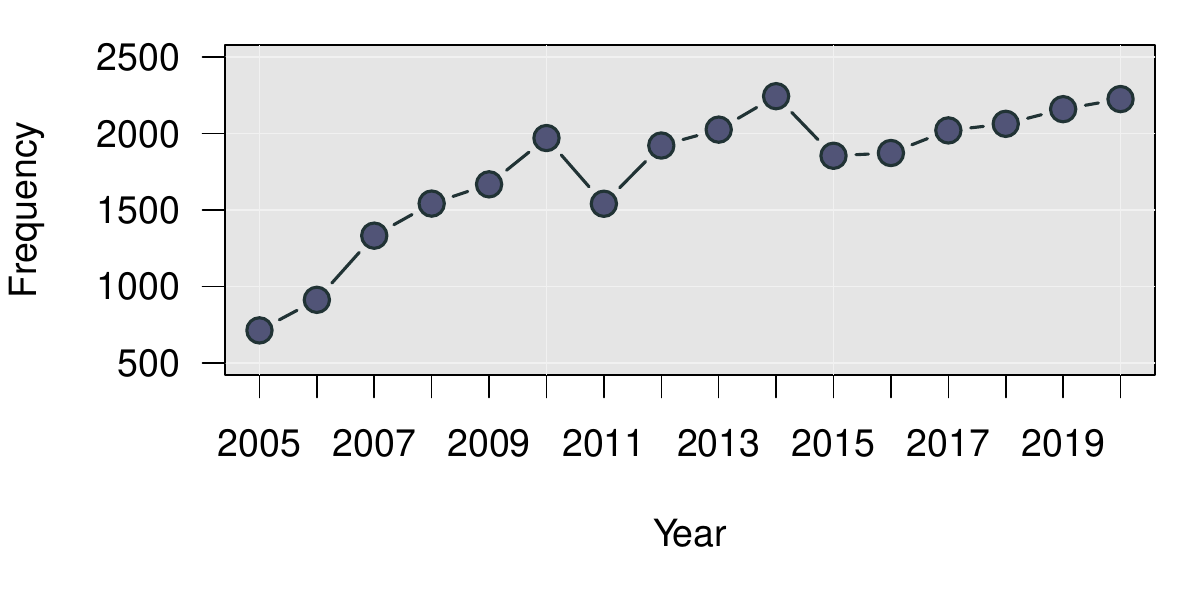}
\caption{Annual Notifications (excluding 2021)}
\label{fig: years}
\end{figure}

As can be concluded from Fig.~\ref{fig: years}, annual submission amounts have
been relatively stable from the 2010s onward. After accelerating growth in the
2000s, roughly about two thousand entries were filed to RAPIX each year. It is
difficult to interpret these magnitudes, but given the size of the EU's internal
market and the amount of consumer products circulating within and across it, a
couple of thousand dangerous products per year seems modest.

\section{Review}

The short review covers seven distinct but overlapping themes: the general,
historical background, and the core regulations, administration,
standardization, product categories, notifications, recalls, and liability,
respectively.

\subsection{Background}

Product safety is present in the Treaty on the Functioning of the European
(TFEU). Like in many policy domains, the legal basis builds upon the functioning
of the internal market for the free functioning of goods, people, services, and
capital. In addition to this general clause specified in the TFEU's Article 26,
the member states have agreed upon the prohibition of import and export
restrictions among themselves with Articles 34 and 35. However, the subsequent
Article 36 states that indiscriminating restrictions are possible for goods on
the grounds of protecting the health and life of humans, animals, or plants,
among other things. Furthermore, Article 12 states a general goal of consumer
protection when legislating and enforcing laws in the European Union, and
Article~191 extends the overall health protection goal toward environmental
considerations. Before the Lisbon Treaty in 2007, which incorporated the
Maastricht Treaty as the TFEU, similar protective clauses were specified in the
Treaty Establishing the European Community. As the European Economic Community
was turning to the European Union, tighter harmonization was required also for
product and food safety.

Instead of focusing on specific products, common European legislative framework
was sought on four strategic areas: fair trading, public health, public
controls, and consumer information, unified by
standardization~\cite{CampbellPlatt94, Howells00}. These strategic goals were
based on the so-called New Approach to regulation, which has generally been
perceived as a success and thus an important factor for the European
integration. In terms of jurisprudence, it started as a response to a seminal
1985 case in the Court of Justice of the European Union (CJEU). The decision
reached by the court was significant for two reasons. Both were related to the
internal market. On one hand, it established the so-called principle of mutual
recognition (i.e., goods sold in one member state must have access to the whole
internal market); on the other, it mandated a set of essential public interest
safety requirements for products~\cite{Ullrich19}. The New Approach led to
Directive 92/59/EEC on general product safety. It is based on five general
principles, as follows:
\begin{enumerate}
\item{The directive's \textit{scope} covers products intended for consumers,
  including new, used, or reconditioned products but excluding second-hand
  products. Safe products, in turn, refer to products that under normal, or
  reasonably foreseeable conditions of use, pose only a minimal risk. The
  antonym is a dangerous product. When assessing a risk, packing, instructions,
  and related factors should be taken into account in addition to a given
  product's characteristics in itself.}
\item{The general safety \textit{requirements} obligate producers to place only
  safe products on the internal market. While these are specified either by
  European standards or, in the absence of such standards, national laws enacted
  in the member states, compliance relies strongly on industry self-regulation
  and accreditation. In addition to these safety requirements, producers must
  provide adequate information to consumers and ensure that identification of
  individual products and product patches is possible after their release to the
  internal market.}
\item{The member states are obliged to ensure compliance through properly
  authorized national \textit{authorities}. Their obligations range from
  compliance monitoring and safety checks to \textit{ex~ante} prohibitions for
  market entry and \textit{ex post} withdrawal of products.}
\item{The member states are further mandated to provide \textit{notifications} to the European Commission about any measures taken regarding dangerous products.}
\item{The \textit{Commission} is empowered to inform other member states in case
  a given member state undertakes an emergency action for dangerous
  products. If a EU-wide solution is required, the Commission has also a right
  to enforce a withdrawal of a dangerous product. Finally, common EU
  institutions coordinate product safety issues between the member states.}
\end{enumerate}

The Maastricht treaty prompted an update to the product safety directive. In
particular, the TFEU's Article 169 strengthened the legal basis for consumer
protection, including on health and safety issues. The resulting policy-making
in the 1990s led to Directive 2001/95/EC, also known as the general product
safety directive (GPSD). It is the directive in force today. Although the
directive made many amendments and clarifications to the 1990s one, the five
general principles remained largely unchanged. Among the amendments and
clarifications are obligations for supply chain distributors of products. The
GPSD also substantially extended the notification framework and information
exchange procedures with the RAPEX architecture. Despite the architecture,
further alterations were required for more efficient monitoring of the internal
market. To this end, the 1990s Regulation (EEC) 339/93 was repealed with
Regulation 768/2008 (hereafter, MSR) for more rigorous market surveillance of
dangerous products. In general, the MSR strengthens the GPSD. A particular
emphasis is placed on national accreditation authorities, serious risks, and
further information exchange provisions.

Today, the GPSD and the MSR are the main effective legislations. That said,
reforms were attempted throughout the 2010s, largely due to the emergence of
electronic commerce. Already in 2013 a new regulatory package was attempted, but
it got struck in a legislative limbo, as did the results from a 2016 evaluation
by the Commission~\cite{Ullrich19}. In practice, only the mutual recognition
principle has been clarified with Regulation 764/2008 and its successor,
Regulation (EU) 2019/515. Both had only a minimal impact upon the existing
product safety legislations. A new consultation for product safety was launched
in 2020 in a conjunction with larger planned reforms on electronic commerce,
digitalization, and related aspects affecting the single market. Even though it
is too early to evaluate the impact from the feedback, it is worth remarking
that the political trenches were dug as could be expected. Regarding electronic
commerce, platform companies, such as Ebay~\cite{ebay20}, argued that platforms
should be exempted from tighter constraints, while consumer and civil society
groups, such as~BEUC~\cite{beuc20}, pointed out that a substantial amount of
products purchased via platforms were already non-compliant with the EU laws and
technical standards.

\subsection{Administration}

The administrative framework is typical to the European Union in general;
administration is decentralized to the member states. Within the safety domain,
however, the decentralized framework is an exception because many other sectors
(such as pharmaceuticals, transport, and aviation) are primarily administrated
through specific EU agencies. In contrast, for product safety the EU-level is
generally reserved for coordination and information exchanges. In fact, none of
the articles in the TFEU establish a particular requirement for a EU-level
competency. Instead, the legal basis for pan-European administration of product
safety has largely been justified with the harmonization measures specified in
the TFEU's Article~114~\cite{Herzmann15, Purnhagen13}. In practice, these
measures include standardization and information exchanges. These were specified
also in the GPSD and the MSR alongside the enforcement at the national level
(Articles 6--10 in the former and Articles 2 and 16 in the latter). As the
administration is typical to the EU, so are the impediments and problems.

\begin{figure*}[th!b]
\centering
\includegraphics[width=\linewidth, height=4.0cm]{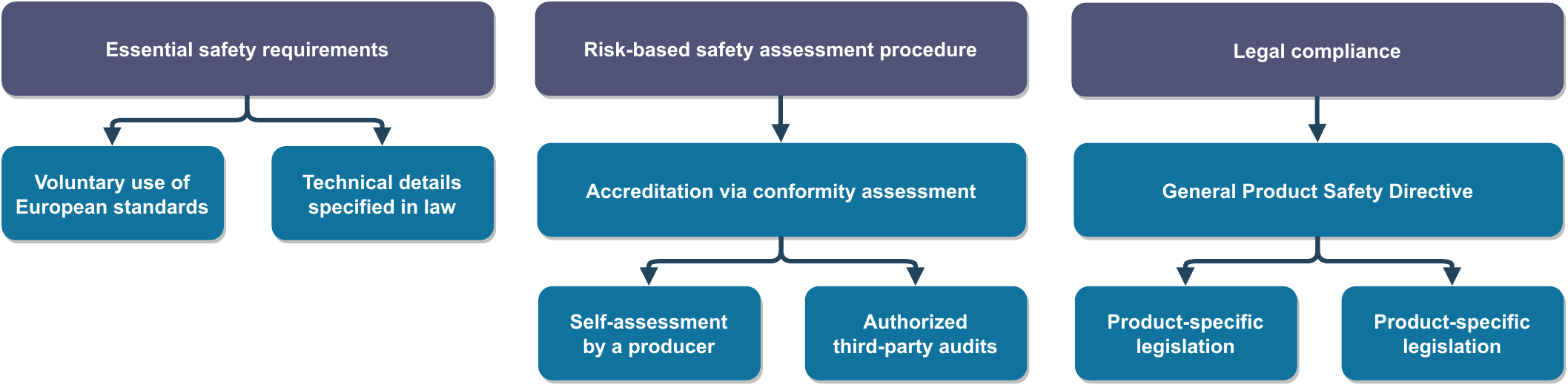}
\caption{The General Logic of the Safety Policy in the European Union (adopted
  from \cite{EC20b} with alterations)}
\label{fig: safety}
\end{figure*}

To some degree, decentralized administration at the national level has
maintained the historical cross-country differences and thus fragmentation
across Europe. Depending on a study, three, four, or five different consumer
protection regimes can be identified in Europe. According to one classification,
there has been a Nordic \textit{negotiation model} (industry associations and
individual companies negotiate directly with consumer associations for common
policy goals), a \textit{protection model} with France as an example (consumer
associations and the state have had a strong influence upon policy goals), and
an \textit{information model} with Austria and Germany as examples (industry
associations and the state have had a dominant role)~\cite{Austgulen20}. In
addition, it is possible to identify further models by casting the focus on the
British-influenced administrative tradition as well as the eastern and southern
member states.

These regimes are not unique to product safety administration and consumer
protection in general. For instance, a comparable administrative
system---including the notification mechanism---has been established for
European cyber security~\cite{Ruohonen16GIQ}. Also privacy and data protection
in the EU share many similarities in terms of administration and
coordination---as well as the associated problems~\cite{Ruohonen20IS}. These
problems include a lack of resources, funding, and expertise in some countries,
generally inconsistent enforcement, poor testing facilities in some countries,
powerlessness in terms of sanctions, diverging legal interpretations, and
general fragmentation~\cite{Hodges04, Klaschka17, Ullrich19,
  vanAken02}. Given the Commission's limited power---it cannot even act as an
arbiter in case the member states disagree on product safety issues, many of the
problems likely prevail in the foreseeable future. All this said, there has been
a high degree of coherence for some consumer products due to standardization.

\subsection{Standardization}

The EU legislations for product safety rely strongly standardization. According
to the New Approach, legislative harmonization establishes essential
requirements required for an entry to the internal market, but discretion is
allowed for producers regarding the technical standards that fulfill these
requirements~\cite{Cordero09}. The role of standards in the safety policy's
general logic is illustrated in Fig.~\ref{fig: safety}.

The difficulties for covering new products through regulation explain the
difference between legislation and standardization. But the difference can be
interpreted also as a way to separate politics from technical expertise; or,
rather, to balance parliamentary legislation procedures with industry
self-regulation~\cite{Dewitt99, Joerges10}. This balancing has also raised
questions about the dominance of the latter over
accountability~\cite{Hodges04}. In terms of legal wordplays, according to the
GPSD, a product is \textit{deemed} safe whenever it complies with a given
European or national legislation. Yet, according to Articles 3 and 4 in the
GPSD, a product is \textit{presumed} safe insofar as it conforms to voluntary
national standards transporting harmonized European standards. Here, European
standards refer to those drafted by European standardization bodies in
accordance with Directive (EU) 2015/153. It includes technical specifications
that specify quality, performance, testing, safety, packaging, and related
dimensions, but excludes radio and television broadcasting, telecommunications,
and financial services, among other things. The major European standardization
bodies for harmonized standards are the European Committee for Standardization
(CEN), the European Telecommunication Standards Institute (ETSI), and the
European Committee for Electrotechnical Standardization (CENELEC). All three are
linked to international standardization organizations and their committees. From
the 2000s onward, European standardization efforts have generally leaned toward
industry consortia and co-regulation in order to improve competitiveness and
innovation~\cite{Blind08, Howells14}. Safety standardization is not an
exception.

In some sectors, there has been some confusion between compliance to European
standards and compliance to European safety legislations~\cite{Mak17}. In many
sectors, however, compliance questions are relatively straightforward for
producers due to the work done by the European standardization organizations who
translate the essential requirements into technical specifications. The CEN, in
particular, has released thousands of technical standards over the decades. With
these standards, accreditation provides the presumed safety conformity
requirement~\cite{Dewitt99}. The prime example is the CE marking established
with Directive~93/68/EEC, which was later augmented with the Commission's
Decision~768/2008/EC. It is not a certification but a declaration of conformance
and due diligence. Typically, the CE marking is obtained by self-certification
and following of harmonized European standards, with or without additional
assessments by a national safety agency or third-party
auditors~\cite{Cordero09, Playle11}. However, the procedure has been an
exception rather than the rule for consumer products.

Fragmentation is present also in terms of standardization. Many complex products
need to comply with multiple legislations \textit{and} multiple standards. A
further problem has been the GPSD's generality; Article~3 implies that the
presumed safety assumption rests on national standardization, which is required
to transpose the various European standards. It is thus no wonder that the three
European standardization organizations have called for unified EU-level
standards, which, according to their position, are achievable via better
coordination, funding, and strategic thinking~\cite{CENCENELEC20,
  ETSI20}. Finally, it is important to underline that compliance with European
standards and their national transpositions is voluntary. While complying with
these provides the presumed safety, it is still possible to place products that
are only deemed safe to the internal market. If the products turn out to be
unsafe, withdrawal of these may follow by national authorities, as soon
discussed.

\subsection{Product Categories}

The New Approach pushed the regulatory work toward general product categories
and essential safety requirements for these. A few notable legislations for
product categories are enumerated in Table~\ref{tab: sectoral}. These are the
directives to which amendments were made with the CE-marking
directive. Of these, the old voltage Directive 72/23/EEC is particularly
noteworthy. It was this directive after which the New Approach was largely
modeled.

\begin{table*}[th!b]
\centering
\caption{Notable Safety Directives for Specific Products}
\label{tab: sectoral}
\begin{threeparttable}
\begin{small}
\begin{tabular}{lll}
\toprule
Sector & Earliest legislation & Latest legislation \\
\hline
Burning gaseous fuels & Directive 90/396/EEC & Directive 2009/142/EC \\
Construction products & Directive 89/106/EEC & Regulation No 305/2011 \\
Electrical equipment (voltages) & Directive 72/23/EEC & Directive 2014/35/EU \\
Electromagnetic compatibility & Directive 89/336/EEC & Directive 2004/108/EC \\
Hot-water boilers (fueling) & Directive 92/42/EEC & -- \\
Implantable medical devices & Directive 90/385/EEC & -- \\
Machinery & Directive 89/392/EEC & Directive 2006/42/EC \\
Non-automatic weighing instruments & Directive 90/384/EEC & Directive 2014/31/EU \\
Personal protective equipment & Directive 89/686/EEC & Regulation (EU) 2016/425 \\
Simple pressure vehicles & Directive 87/404/EEC & Directive 2014/29/EU \\
Telecommunications terminal equipment & Directive 91/263/EEC & Directive 93/68/EEC \\
Toys & Directive 88/378/EEC & Directive 2009/48/EC \\
\bottomrule
\end{tabular}
\end{small}
\end{threeparttable}
\end{table*}

Even with this small snapshot of the specific EU legislations, it can be
concluded that the scope is wide; from machinery and vehicles to medical devices
and toys. Furthermore, many products (such as machinery or motor vehicles) must
comply with multiple legislations~\cite{Playle11}. It is also worth remarking
that some notable consumer products, such as cosmetics~\cite{Pauwels10}, are
missing from the listing. A~further important point is that the relation of the
specific legislations to the GPSD and its predecessor has been a source of some
confusion~\cite{Howells14}. In principle, the GPSD and associated national laws
should apply in case there are no specific legislations for a given product
category. The MSR is also explicitly specified as \textit{lex spesialis}; it is
applicable only insofar as there are no specific legislations for market
surveillance, as is the case with drug precursors, medical products, vehicles,
and aviation. As the GPSD also states that the risks specified in it are
applicable unless overruled by a specific legislation, it should be taken into
account alongside any existing specific legislations.

\begin{figure}[th!b]
\centering
\includegraphics[width=8cm, height=8cm]{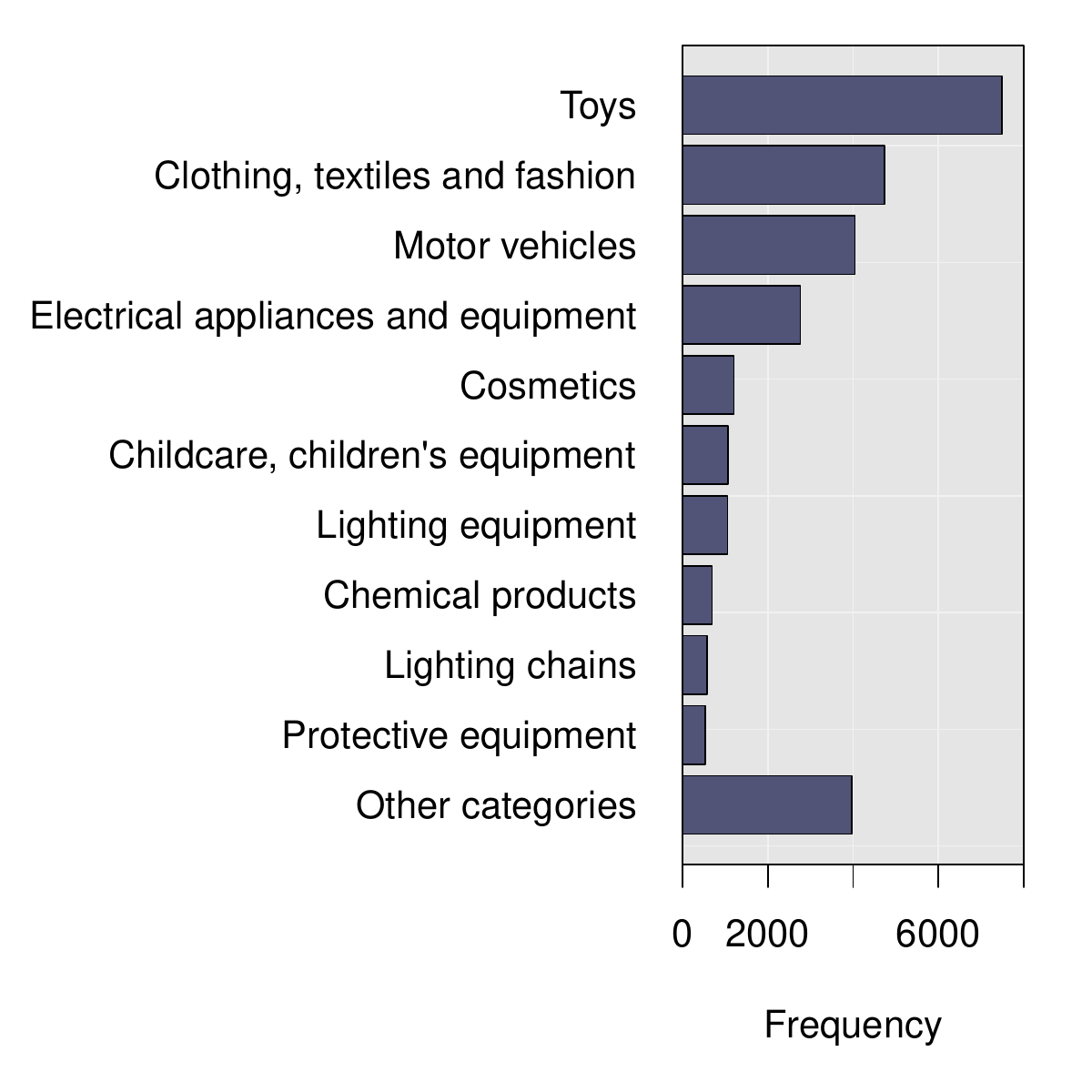}
\caption{Top-10 Product Categories}
\label{fig: categories}
\end{figure}

To get a sense on the actual, realized safety risks across the categories,
Fig.~\ref{fig: categories} shows the ten most frequent product categories
(alongside a catch-call group for other categories) across all entries filed to
the RAPEX, from 2005 to mid-January 2021. Toys, clothing and textiles, motor
vehicles, and electronic devices constitute the majority of reported unsafe
consumer products. Together these categories account for about 68\% of all RAPEX
filings. Particularly the large amount of dangerous toy products is interesting
and surprising. It is, however, difficult to speculate the reason for this
result; one explanation could be that the safety of toys is vigorously enforced
by the national safety authorities. This would align with the long tradition of
considering children as a particularly vulnerable consumer
group~\cite{Mugica90}. Numerous European standards have also been specified for
toys~\cite{Mak17}. In addition to toys, many filings were made about clothing,
textiles, and fashion items, motor vehicles, and electrical appliances.

\begin{figure}[th!b]
\centering
\includegraphics[width=8cm, height=10cm]{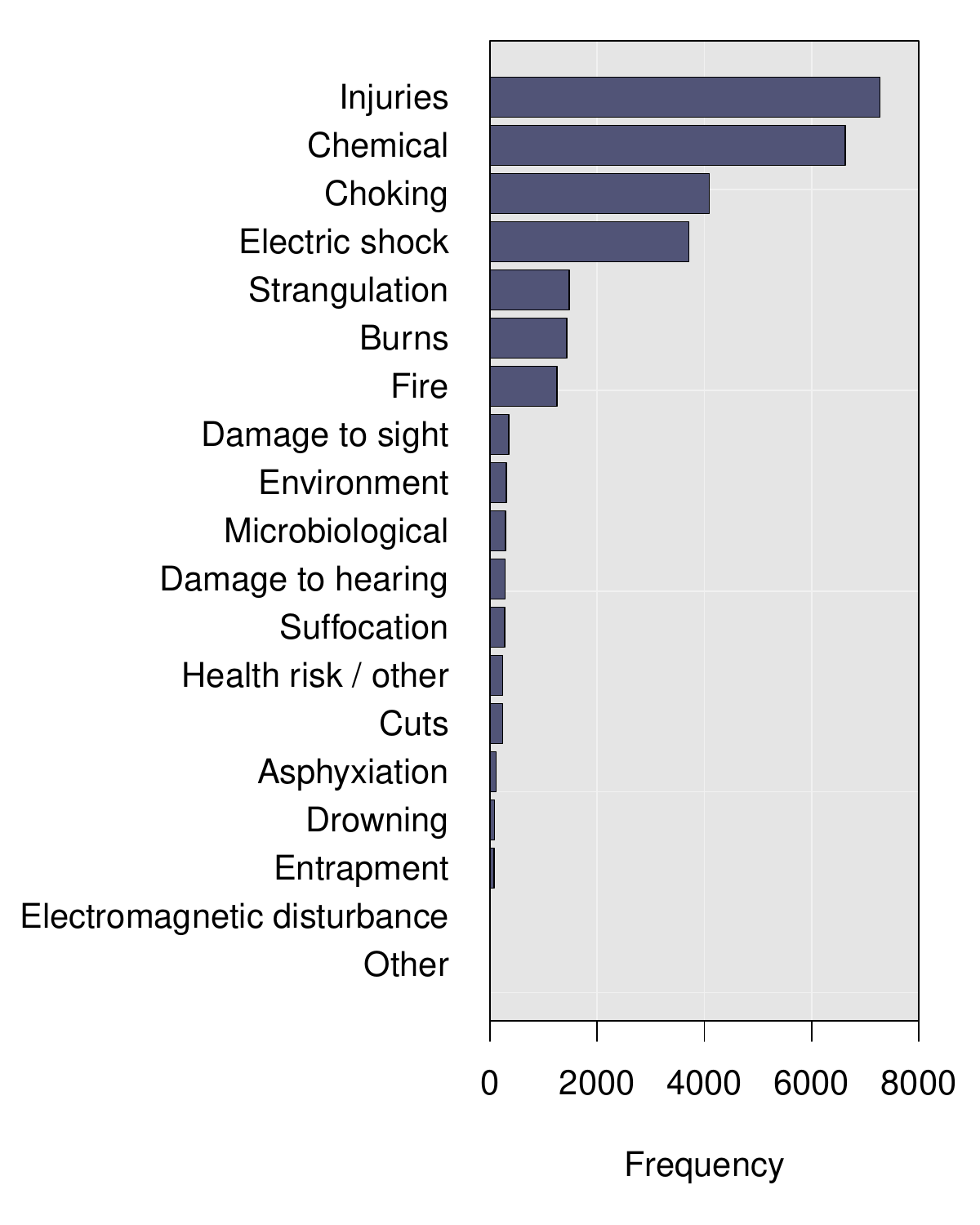}
\caption{Risk Types}
\label{fig: risk types}
\end{figure}

\begin{figure}[th!b]
\centering
\includegraphics[width=8cm, height=8cm]{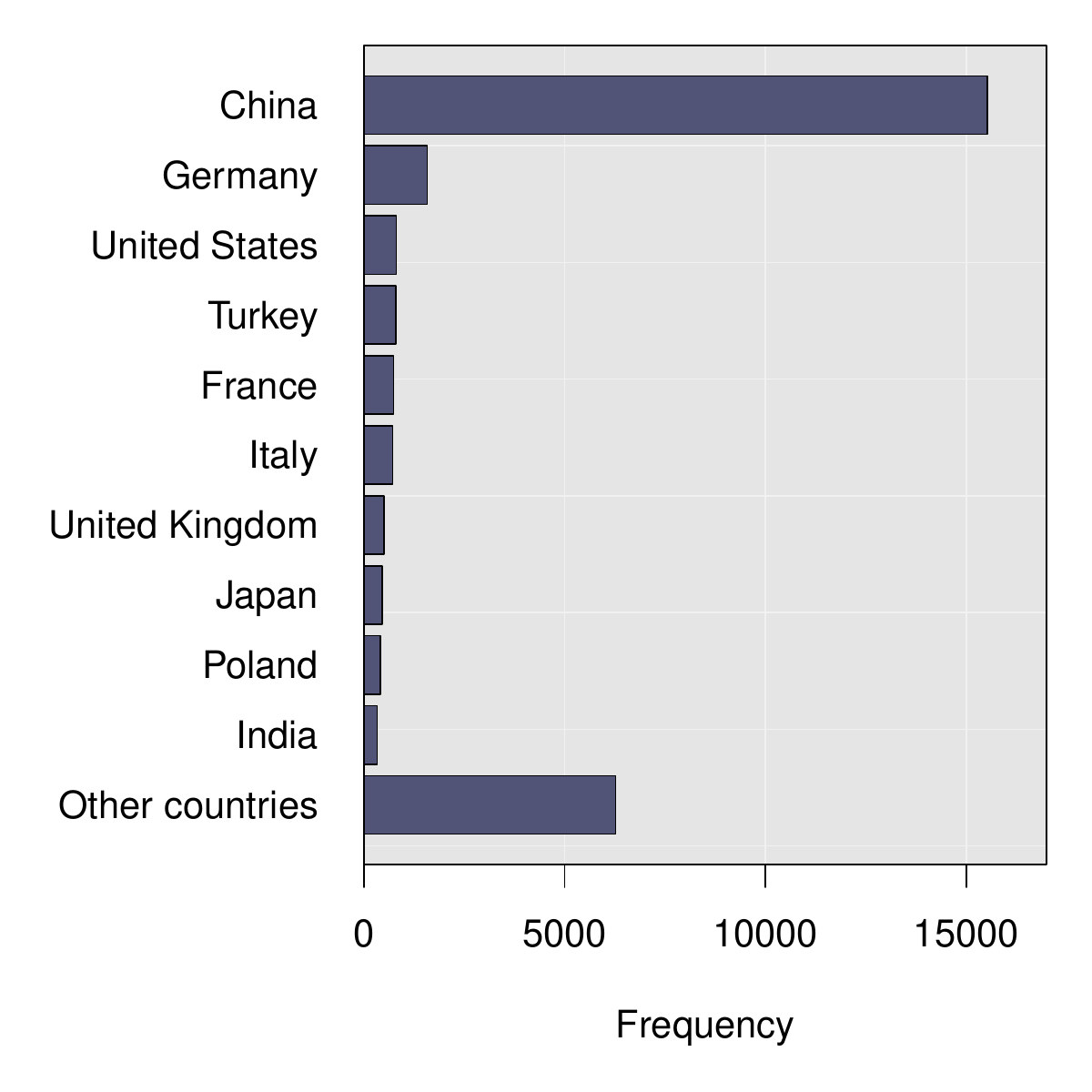}
\caption{Top-10 Countries of Origins}
\label{fig: imports}
\end{figure}

Three additional points are worth making about the RAPEX entries. First: besides
potential variance in terms of enforcement practices, these observations likely
reflect the large amount of consumer products in falling to these
categories. Second: the product categories further reflect the typical risk
types shown in Fig.~\ref{fig: risk types}. For instance, about 34\% and 32\% of
dangerous toy products had caused chemical and chocking injuries,
respectively. From all products with injury risks, about 41\% were toys, 16\%
cosmetics, and 16\% clothing, textiles, and fashion items. Last: the frequencies
in Fig.~\ref{fig: categories} correlate with the origin from which the products
were imported to the internal market. As can be seen from Fig.~\ref{fig:
  imports}, the clear majority of the dangerous products were imported from
China. For instance, as much as 78\% of the dangerous toy products were
manufactured in China. As such, the observation cannot be strictly interpreted
to imply that Chinese products would be particularly risky, given that most toy
products are manufactured in China. Nevertheless---given that about 19\% of
total imports to Europe were from China in 2020~\cite{Eurostat20a}, it seems
sensible to relate the observation to the increasingly complex supply chains for
consumer and other products.

\subsection{Notifications}

The notification mandates set in Directive 92/59/EEC were modeled after existing
procedures laid down for some other sectors. Notably, notification procedures
for pharmaceuticals were laid down already in the mid-1970s and early 1980s with
Directives 75/319/EEC and 81/851/EEC. These were followed by Directives
82/894/EEC and 89/662/EEC concerning animal diseases and products of animal
origin, respectively. Furthermore, a year after the Chernobyl disaster,
Council's Decision 87/600/Euratom had established a system for rapid exchange of
information in radiological emergencies. The GPSD generalized these procedures
toward a general requirement: according to Article~5, both producers and
distributors must inform a national authority whenever they know that a product
contains safety risks, and coordinate with them on any preventive measures
taken. In practice, the mandate largely rests on producers' own risk assessment
and management procedures~\cite{Howells14}. With accreditation, however, the
national accreditation authorities should inform each others according to the
MSR's Article~12. It is also worth mentioning notification requirements between
the national authorities and the Commission, as well as information exchanges in
situations requiring rapid interventions, as specified in the GPSD's
Article~11. Finally, there is the notification requirement toward the subjects
being protected, the consumers; as specified in Article~16, information about
safety risks should be available to the public according to transparency and
other good administration practices. Some criticism has been leveraged about
passivity in this informing obligation~\cite{Herzmann15}. At the EU-level this
criticism finds its target in the RAPEX infrastructure, as well as in its use by
national authorities and media.

\subsection{Recalls}\label{subsec: recalls}

Product withdrawals were an important element already in the legacy
Directive~92/59/EEC. Both the GPSD and the MSR clarified the powers granted to
national authorities in this regard. The MSR's Article~21 and the GPSD's
Article~5 stipulate that national authorities have the right to ban new market
entries as well as withdraw existing dangerous products from the market. This
withdrawal deterrence has provided an important incentive for producers to
ensure safety of their products~\cite{Cavaliere04}. Though, Article~5 in the
GPSD somewhat loosens the obligations for producers by emphasizing that recalls
should only be used as a last resort. This concession is understandable because
recalls have been a controversial issue for both producers and regulators: for
the former---besides plain economic losses, these may interfere with insurance
schemes; for the latter, there has been a fear that producers will externalize
difficult recall decisions to them~\cite{Howells14}. In terms of product safety,
however, withdrawals are particularly important as these concern products
consumers are already using.

\begin{figure}[th!b]
\centering
\includegraphics[width=8cm, height=8cm]{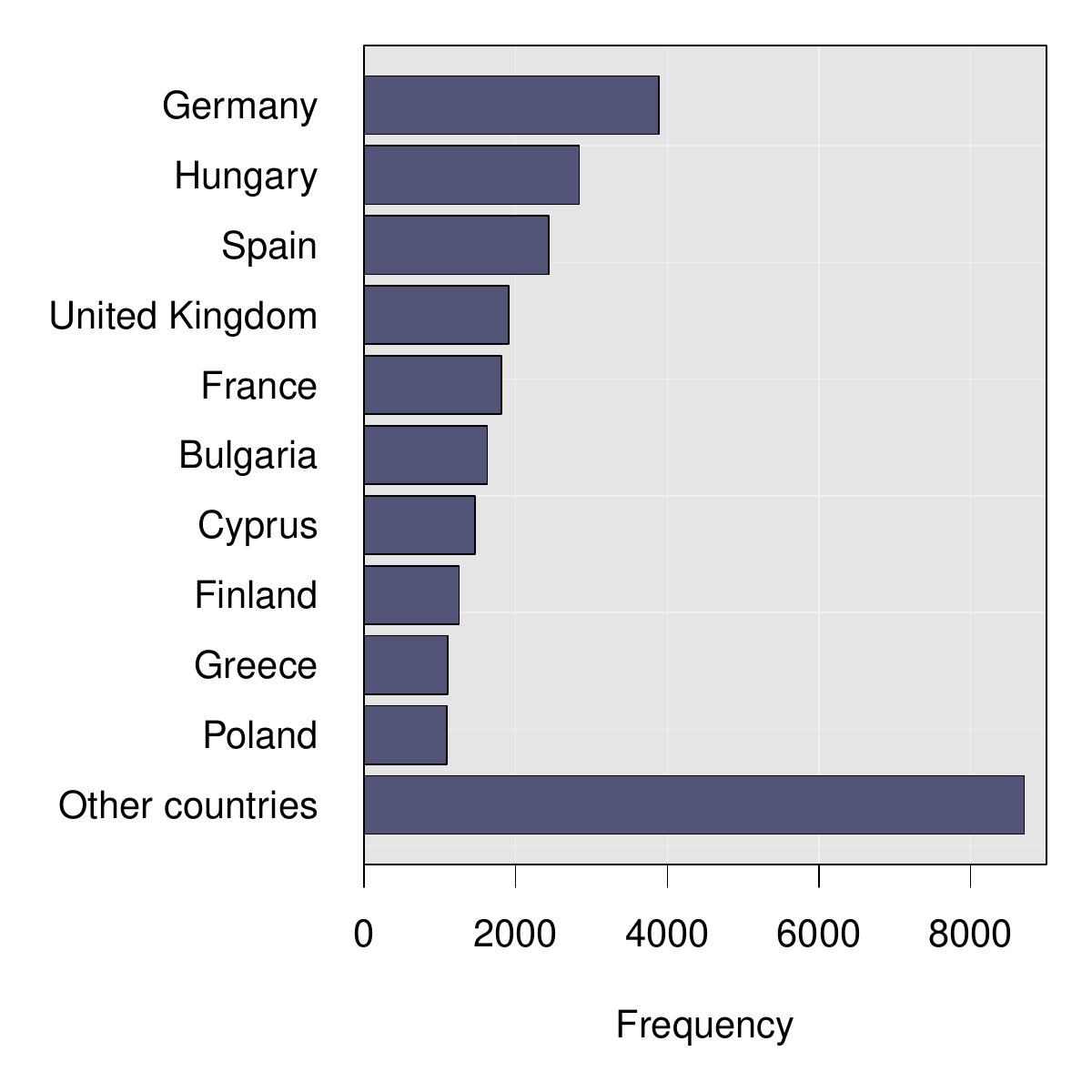}
\caption{Top-10 Country Filings to RAPEX}
\label{fig: submission}
\end{figure}

Recalls have also been frequent according to the empirical RAPEX sample. By
using simple regular expression searches, about 35\% and 33\% of the products'
descriptions contain the words \textit{recall} and \textit{withdraw},
respectively. Other keywords are far less common; these include \textit{ban}
(13\%), \textit{reject} (6\%), \textit{correct} (5\%), and
\textit{destruct} (3\%), among a few other auxiliary terms. Therefore, it can be
concluded that the national safety authorities have also actively used their
product recall powers. This observation reflects the many individual
high-profile withdrawal cases with serious economic
implications~\cite{Rausand09}. But the RAPEX does not, unfortunately, contain
enough information to deduce whether the many product withdrawals undertaken
were done after a proactive notification from a producer or a distributor. Nor
is it sufficient to deduce about the criticism that some countries have
preferred voluntary corrective actions~\cite{vanAken02}. If filings to the RAPEX
is used as a proxy, a large cross-country variance is indeed present. For
instance, more filings were made from Hungary than from France and the United
Kingdom (see Fig.~\ref{fig: submission}). This observation reinforces the
earlier points about general fragmentation and incoherence. A further relevant
question is whether the relatively low share of market entry bans indicates
improvement potential for more proactive \textit{ex~ante}~enforcement.

\subsection{Liability}

A final important aspect of the New Approach was the adoption of the product
liability Directive 85/374/EEC. It extends the hybrid approach of risk-based
assessments, harmonization and standardization, and market surveillance with
explicit allowance of litigation. Safety is explicitly spelled in the
directive's Article~6, and legal basis for penalties is possible by translating
the ``polluter pays principle'' implicitly present in the TFEU's Article~191 to
the safety context~\cite{Purnhagen13}. The directive introduced strict
liability for producers, but there are several means by which a producer may
argue against a claimant: improper use of the product in question, the presence
of a defect at the time when the product was launched or marketed, the lack of
scientific knowledge at the time when the defect was noticed, the presence of
the defect despite compliance with legislations and standards, and so
forth~\cite{Raposo15}. For these reasons, Directive 85/374/EEC cannot be fully
interpreted to be a strict liability law---in practice, the necessity to
demonstrate the damage, the defect, and the causal relation between them make
the directive negligence liability law instead~\cite{Faure03, Zech21}. It has
nevertheless provided an incentive for producers to ensure
safety~\cite{Rausand09}. All in all, it is important to stress that a
``regulation through litigation'' approach has generally been limited in the
safety domain; reforms have been pushed forward by the organized interests of
producers, consumers, and experts rather than by litigation, which is
historically paradoxical due to the CJEU's central role behind the New
Approach~\cite{Purnhagen13}. Although hundreds of cases have been brought to
courts throughout Europe~\cite{EC18a}, high-profile cases affecting large
consumer groups have been rare for consumer goods.

There is no single explanation why the litigation-centric approach did not fully
emerge. In addition to lobbying, the reasons include: the experiences in the
United States, where strict liability was established for product safety in the
1960s but which came under criticism in the 1980s for its alleged economic
impacts, the state-of-the-art defense provisions in Directive 85/374/EEC, the
European tradition of compensations through welfare states, underdevelopment of
insurance schemes for producers, a lack of contractual relationships in some
cases, litigation costs for consumers, and the recall deterrence, among other
things~\cite{Bradgate02, Cavaliere04, EC18a}. History offers another plausible
explanation: Directive 85/374/EEC was enacted before there was any talk about
the subsidiarity principle, before the Maastricht
Treaty~\cite{Faure03}. Regarding more recent times, an alternative
interpretation would be that the GPSD, the auxiliary product-specific
legislations, and European standards have worked efficiently in preventing
dangerous goods from entering the internal markets to begin with. However, not
all products are protected from safety threats equally. Although the question
has been more about security than safety---a distinction, which, as noted, is
not always clear, software liability has not generally progressed despite a long
global debate~\cite{Moore13, Ryan03}. Europe is no
exception~\cite{Cabral20}. As soon discussed, software and information
technology products and services have also introduced other major security and
safety challenges.

\section{Conclusion}

This review addressed consumer product safety regulations in the European
Union. Although the topic is complex---covering tens if not hundreds of
legislations, many of which address highly specialized products and science
associated with these, some general points can be briefly made about the safety
regulations. These are largely based on a hybrid approach. Thus, first, the GPSD
and the MSR provide the umbrella legislations augmented with product-specific
legislations and European standards. Both are relevant for compliance. Second,
however, the hybrid policy is generally based on the precautionary
principle. Although actual practice may be different, risk analysis carries a
particular weight in the overall policy. Third, the administration is
decentralized to the member states. Like in many related policy domains, the
EU-level is mainly reserved for coordination and information exchanges. Fourth,
even though possible, litigation has been relatively rare; product recalls
remain the main deterrence.

Although recommending change for the sake of change is one of the deadly sins of
policy analysis~\cite{Meltsner86}, there are many indicators of a reform
need. Many of these have also been acknowledged in the EU. The Commission's
recent summary report~\cite{EC20a} of the feedback on the GPSD suffice to
outline the main factors behind the need. These are:
\begin{itemize}
\item{Although the Commission pointed out \textit{inconsistency} only in terms
  of food-imitating products, incoherence and fragmentation are a problem
  for the whole product safety policy. As is common in the EU, the reason is
  partially explained by both vertical (national policies versus EU policies)
  and horizontal (variance across the member states) incoherence. The issue goes
  beyond the safety legislations. For instance, the product liability directive
  has not been uniformly adopted in Europe~\cite{EC18a}. Another notable aspect
  relates to European standards, which still need to be transposed to national
  standards. The position of European standardization organizations seems rather
  unequivocal: there is a need for unified standards at the
  EU-level. Incoherence and fragmentation are not the only reasons; industry
  competitiveness is also a concern with standards. According to
  interviews~\cite{Bujok17}, in some sectors particularly small and
  medium-sided enterprises have faced challenges in selecting appropriate
  standards and implementing these for their products.}
\item{\textit{Enforcement} has been a closely related problem. Although the
  situation cannot be described as a race to the bottom whereby some member
  states would deliberately weaken their enforcement
  responsibilities~\cite{Faure03}, resources, expertise, legal interpretations,
  and incentives, among other factors, vary across the member states. As the
  Commission noted, the effectiveness of recalls is a particular concern because
  many consumers continue to use dangerous products despite recall notices and
  withdrawals of product stocks in sale. As for potential solutions, as the
  European Parliament emphasized in its 2019 resolution~\cite{EP19a}, further
  harmonization, sufficient resourcing, and better coordination may improve the
  situation. As there are EU agencies for safety in some sectors, it is also
  worth contemplating whether or not a move toward EU-level administration might
  offer a longer term solution.}
\item{\textit{Market surveillance} has been a third common problem. Closely
  related to the enforcement problems, according to the Commission, there has
  been a lack of tools and instruments to impose effective sanctions. As with
  security, the complex global supply chains constitute a major problem for
  product safety in Europe and elsewhere. Traceability, product integrity, and
  supply chain management are increasingly important for safety risk management
  and efficient recalls~\cite{Marucheck11}. Analogous points apply to software
  products and their security~\cite{MITRE21}. Although consumer products may be
  an exception, it can be generally argued that European approaches are not
  sufficient alone; the management of large-scale hazards requires global
  solutions, and with such solutions, political obstacles should never be
  undermined \cite{Posaner21}. Furthermore, within Europe, the enactment of the
  MSR has led to uneven requirements for different products; the incoherence
  across products and across Europe applies also to post-market activities. The
  same applies to RAPEX risk assessment guidelines, which are overly general and
  mainly left to national authorities~\cite{OECD16a}. The general fragmentation
  is again visible.}
\item{\textit{Online platforms} are an increasing problem for product safety and
  its enforcement. As such, the issue is hardly new. The difficulty to identify
  sellers and hold them accountable on electronic commerce marketplaces was
  recognized already in the early 2000s~\cite{Bradgate02}. Yet, only recently
  have the European safety authorities cast their attention to the issue. As an
  example: in total, from 2018 onward, $244$ submission to the RAPEX have
  mentioned either eBay, Amazon, or both. Although the amount is tiny compared
  to the amount of consumer products delivered via these platforms, it is still
  sufficient to conclude that the problem is already recognized also on the side
  of enforcement. A particular problem with these new platforms relates to their
  business model; they act as intermediaries for third-party sellers whose
  responsibilities---or even identities---are not always clear. Like with supply
  chains, the issue is also largely global, and, unfortunately, existing
  international arrangements are soft guidelines at best in this regard
  \cite{Benohr20}. As consumers increasingly purchase products from sellers
  located outside of the EU, ensuring safety has become more difficult for
  European authorities. Many of these products do not enter to the internal
  market as large batches, which complicates the work at customs. In these
  cases, alerts are possible but recalls are difficult.}
\item{\textit{New technologies} are the final problem recognized by the
  Commission---and for a good reason. Technologies such as artificial
  intelligence have fueled the whole safety debate to a new level throughout the
  world. But according to the Commission's auxiliary report~\cite{EC20b}, the
  safety provisions for artificial intelligence applications, robotics, and
  Internet-of-things devices are seen attainable mainly through transparency,
  accountability, and unbiasedness of algorithms, fallback mechanisms, and
  keeping a human in the loop. Although these are sensible goals as such, it
  remains disputed how these could be legislated with sufficient
  rigor---especially when considering that safety science is not quite there
  yet~\cite{Houben21}. For these and other reasons, liability for new
  technologies has been actively debated recently. Given that reforms are
  already underway at this front, it suffices to generally pinpoint the many
  challenges: unclear definitions, including the legal status of artificial
  intelligence systems~\cite{Gudkov20}, different legal notions of liability,
  potential (compulsory) insurance schemes, impacts upon economy and innovation,
  supply chains, and so on~\cite{Bugra20, Cabral20, Zech21}. Another related
  point is the ethical use of artificial intelligence promoted by the EU. Here,
  it remains unclear whether ethics can---or should---be conflated with safety,
  which should be arguably always guaranteed regardless of what is seen as right
  or wrong.}
\end{itemize}

Three additional points deserve a brief discussion. The first point is
practical: the RAPEX system could benefit from more meta-data. For instance, the
textual descriptions often lack sufficient information to deduce about which
legislations or standards were specifically violated. Although some national
authorities have occasionally used general phrasings such as ``machine
directive'', most of the entries lack even such elementary cues. Some criticism
has also been expressed previously regarding missing information about specific
dangerous compounds in some products, and whether these ended to RAPEX due to
producers' self-disclosures or due to testing by national
authorities~\cite{Michalek19}. Further small practical improvements are not
difficult to imagine; there are no longer working hyperlinks pointing
particularly to online platforms, and so on. In terms of bigger implementation
challenges, integration with other safety alert systems remain a priority. Given
the amount of consumer products imported from China~(see Fig.~\ref{fig:
  imports}), supply chain integration seems a particularly noteworthy goal. It
may also provide synergies with security requirements for consumer
products. These points call for comparative cross-country research addressing
both technical aspects and policy challenges.

The second point is about the accreditation and its relation to other ongoing
policy changes. Regulation (EU) 2019/881 established cyber security
certification for information technology products. Even when keeping in mind
that certification should not be equated to accreditation, the directive's
certification scheme is interesting because it differs from the New Approach
tradition: it is not based on harmonized standards enacted by the three European
standardization organizations, and, in some cases, certification via it may
overlap with these standards and the product-specific
legislations~\cite{Kohler20}. Besides aligning with security, the certification
scheme largely builds on global soft guidelines augmented by the institutional
European cyber security authority; a need for further harmonization and
potential standardization is evident also in this regard~\cite{Matheu20}. From
a policy perspective, further incoherence may thus follow in the long-run. This
assertion leads to consider the third and final point.

Last, it is sensible to contemplate whether the whole concept of safety should
be defined better to reflect new realities. Many of the established demarcations
seem outdated, both in academia and in practice. For instance, the distinction
between products and services has acknowledgedly become blurry for legal,
business, and safety considerations~\cite{Bjorkdahl20, Cabral20, EC18a,
  EC20b}. Particularly information technology products have long relied on
different services, nowadays often fueled by data. As the issues with global
supply chains demonstrate, it is also increasingly difficult to make clear-cut
distinctions between producers, suppliers, and distributors. When data powers
capitalism, even consumers are producers, suppliers, and distributors
themselves; they produce, supply, and distribute their data for the new means
of production~\cite{Ruohonen21DJ, Torpey20}. Then, there is the safety concept
itself. The relation between security and safety has always been
ambiguous. Privacy as a safety concern has also long been
debated~\cite{Grimmelmann10}. But there is more. Even information itself can
today be seen as a safety issue---if not a hazard. For instance, recent results
indicate a substantial amount of vaccine misinformation (some of which may also
be intentional disinformation) associated with potentially dangerous products
sold on electronic commerce platforms~\cite{Juneja21}. These and other similar
results restate the general platform problem. But they further call to theorize
and reconceptualize safety.

\balance
\bibliographystyle{splncs03}

\end{document}